\documentclass[sigconf]{acmart}

\usepackage[ruled]{algorithm2e}
\usepackage{multirow}
\usepackage{subfigure}
\usepackage{enumitem}
\usepackage{sistyle}
\SIthousandsep{,}
\usepackage{makecell}
\usepackage{multirow}
\usepackage{enumitem}
\usepackage{bbm}

\AtBeginDocument{%
  \providecommand\BibTeX{{%
    \normalfont B\kern-0.5em{\scshape i\kern-0.25em b}\kern-0.8em\TeX}}}

\copyrightyear{2022}
\acmYear{2022}
\setcopyright{acmcopyright}\acmConference[CIKM '22]{Proceedings of the 31st ACM International Conference on Information and Knowledge Management}{October 17--21, 2022}{Atlanta, GA, USA}
\acmBooktitle{Proceedings of the 31st ACM International Conference on Information and Knowledge Management (CIKM '22), October 17--21, 2022, Atlanta, GA, USA}
\acmPrice{15.00}
\acmDOI{10.1145/3511808.3557527}
\acmISBN{978-1-4503-9236-5/22/10}

\makeatletter
\g@addto@macro\normalsize{%
  \abovedisplayskip 2pt plus1pt 
  \belowdisplayskip 1pt plus1pt
  \abovedisplayshortskip  2pt plus1pt%
  \belowdisplayshortskip  1pt plus1pt
}
\makeatother

\begin{document}

\title{A Contrastive Pre-training Approach to Learn Discriminative Autoencoder for Dense Retrieval}

\author{Xinyu Ma}
\affiliation{
 \institution{CAS Key Lab of Network Data Science and Technology, ICT, CAS}
 \institution{University of Chinese Academy of Sciences}
 \city{Beijing}
 \country{China}
}
\email{maxinyu17g@ict.ac.cn}
 
\author{Ruqing Zhang}
\affiliation{
 \institution{CAS Key Lab of Network Data Science and Technology, ICT, CAS}
 \institution{University of Chinese Academy of Sciences}
 \city{Beijing}
 \country{China}
}
\email{zhangruqing@ict.ac.cn}
 
\author{Jiafeng Guo}
\authornote{Jiafeng Guo is the corresponding author.}
\affiliation{
 \institution{CAS Key Lab of Network Data Science and Technology, ICT, CAS}
 \institution{University of Chinese Academy of Sciences}
 \city{Beijing}
 \country{China}
}
\email{guojiafeng@ict.ac.cn}

\author{Yixing Fan}
\affiliation{
 \institution{CAS Key Lab of Network Data Science and Technology, ICT, CAS}
 \institution{University of Chinese Academy of Sciences}
 \city{Beijing}
 \country{China}
}
\email{fanyixing@ict.ac.cn}
 
\author{Xueqi Cheng}
\affiliation{
 \institution{CAS Key Lab of Network Data Science and Technology, ICT, CAS}
 \institution{University of Chinese Academy of Sciences}
 \city{Beijing}
 \country{China}
}
\email{cxq@ict.ac.cn}

\renewcommand{\shortauthors}{Xinyu and Ruqing, et al.}

\begin{abstract}

Dense retrieval (DR) has shown promising results in information retrieval. 
In essence, DR requires high-quality text representations to support effective search in the representation space. 
Recent studies have shown that pre-trained autoencoder-based language models with a weak decoder can provide high-quality text representations, boosting the effectiveness and few-shot ability of DR models.
However, even a weak autoregressive decoder has the bypass effect on the encoder.
More importantly, the discriminative ability of learned representations may be limited since each token is treated equally important in decoding the input texts.
To address the above problems, in this paper, we propose a contrastive pre-training approach to learn a discriminative autoencoder with a lightweight multi-layer perception (MLP) decoder.
The basic idea is to generate word distributions of input text in a non-autoregressive fashion and pull the word distributions of two masked versions of one text close while pushing away from others.
We theoretically show that our contrastive strategy can suppress the common words and highlight the representative words in decoding, leading to discriminative representations.
Empirical results show that our method can significantly outperform the state-of-the-art autoencoder-based language models and other pre-trained models for dense retrieval.

\end{abstract}



\maketitle

\section{Introduction}

Recently, dense retrieval (DR) has achieved great success on many information retrieval (IR) related tasks, such as web search~\cite{Xiong2021ApproximateNN,Zhan2021OptimizingDR}, open-domain Question Answering (QA)~\cite{Karpukhin2020DensePR,Qu2021RocketQAAO} and fact verification~\cite{Chen2022GEREGE}. 
DR models generally employ pre-trained language models as text encoder to obtain dense representations for queries and documents. 
Then, retrieval with simple similarity metrics can be conducted effectively in the representation space. 
Effective search is based on high-quality text representation learning~\cite{lu-etal-2021-seed}.

Despite the effectiveness of BERT-like language models~\cite{Devlin2019BERT} on learning word representations, they are not good at producing text sequence representations~\cite{Reimers2019SentenceBERTSE,Li2020,su2021whitening}.
Recent studies have demonstrated that autoencoder-based language models can significantly advance the effectiveness and few-shot ability of DR models~\cite{lu-etal-2021-seed}.  
The basic idea is to train a weak autoregressive decoder that reconstructs the input text only from the encoder's encodings.  
In this way, the encoder creates a bottleneck to provide high-quality text sequence representations. 
However, even a weak autoregressive decoder has the bypass effect in which the decoder may ignore the representation and predict the next token only based on previous tokens.
More importantly, the decoder treats each token equally important but common words like in, the, and of, are the majority part of the text.
Therefore, the discriminative ability of dense representations may be limited since the representation will focus more on the common words and thus is not differential with other representations.

To address the above problems, in this paper, we propose a contrastive pre-training approach to learn a discriminative autoencoder with a lightweight multi-layer perception (MLP) decoder.
Specifically, rather than reconstructing texts in an autoregressive fashion, the MLP decoder generates word distributions of input texts in a non-autoregressive fashion to avoid the bypass effect.
We then introduce a novel contrastive learning method to pull the word distributions of two masked versions of one text close while pushing away from others.
We theoretically show that our contrastive strategy can suppress the common words and highlight the representative words when decoding, leading to discriminative representations.
Empirical results verified the effectiveness of our proposed discriminative autoencoder over the state-of-the-art autoencoder-based language models and other pre-trained models.

\section{Related work}\label{sec:related}
In this section, we briefly review the recent studies on designing pre-training methods tailored for dense retrieval.
Early practice in this direction is \cite{Chang2020PretrainingTF}, which proposed three pre-training tasks to resemble the downstream passage retrieval in open-domain QA.
Specifically, Inverse Cloze Task (ICT) is a commonly-adopted task, where the basic idea is to predict the context for a randomly sampled sentence from the Wikipedia page. 
The most related work with ours is SEED~\cite{lu-etal-2021-seed}, which pre-trains an autoencoder with a 3-layer weak Transformer decoder while restrict its attention spans. 
Another line of research is to design new model architectures \cite{Gao2021CondenserAP} for dense retrieval. 
Researchers have also investigated to leverage contrastive learning method to learn sequence representations~\cite{Reimers2019SentenceBERTSE,Gao2021SimCSESC,Ma2022COSTA}.
But these methods are not suitable for learning high-quality document 
representations as we shown in Section~\ref{sec:main_results}.


\section{Our Method}

In this section, we describe our Contrastive Pre-training a Discriminative AutoEncoders (CPDAE) for dense retrieval.

\subsection{Model Architecture}

Basically, the model architecture of our CPDAE is composed of a Transformer encoder and a MLP decoder.

\subsubsection{Encoder}
The encoder aims to encode the input text into low-dimensional dense representations.
We use Transformer~\cite{vaswani2017attention} as the encoder.
Given an input text $d_i=([CLS],w_1,...,w_n)$, a special token [CLS] is added to the front of $d_i$ to represent the whole text. 
For each $d_i$, we follow the masking strategy in  BERT~\cite{Devlin2019BERT} to randomly mask its several tokens twice, to obtain two masked versions $d_i = \{d_{i}^{'}, d_{i}^{''}\}$. 
We take the [CLS] representation of the last Transformer layer as the whole text representation,
\begin{equation}\label{encode}
\textbf{h}_{[CLS]} = Encoder(d_i), ~~~ \textbf{h} \in \mathbb{R}^{H},
\end{equation}
where $H$ is the hidden size.

\subsubsection{Decoder}
The MLP decoder is to recover the input text solely from the text sequence representations. 
Specifically, the decoder includes two layers of feed-forward neural network (FFN) with a non-linear activation function Gelu~\cite{Hendrycks2016GaussianEL} and a LayerNorm function~\cite{Ba2016LayerN}.
Then the MLP decoder maps the text representation to word distributions,
\begin{equation}\label{decode}
\textbf{z} = Decoder(\textbf{h}_{[CLS]}), ~~~ \textbf{z} \in \mathbb{R}^{|V|},
\end{equation}
where $V$ is the vocabulary. 

\subsection{Contrastive Pre-training}
Our contrastive pre-training includes three pre-training objectives: reconstruction loss, contrastive loss and masked language modeling (MLM) loss.

\subsubsection{Reconstruction Loss}
Our non-autoregressive reconstruction is to predict which words in the vocabulary appear in the input text by generating a word distribution. 
Specifically, given the prediction vector $\textbf{z}$ in Eq.~(\ref{decode}), we apply the $Sigmoid$ function for each value $z^j$ in $\textbf{z}$ separately to obtain a valid probability, i.e., 
\begin{equation}\label{alg:sigmoid}
\begin{aligned}
\hat{z}^j = Sigmoid(z^j), j=1,...,|V|,
\end{aligned}
\end{equation}
where $\hat{z}^j$ ranges from 0 to 1 and indicates the probability of $j$-th  word in the vocabulary $V$ appearing in the input text. 
The reconstruction loss is formulated as a multi-label classification problem and computed with the cross-entropy function, 
\begin{equation}\label{recs}
\begin{aligned}
\mathcal{L}_{REC} = -{\sum_{j=1}^{|V|}{(y^{j}\log{\hat{z}^{j}}+(1-y^{j})\log{\hat{z}^{j}})}}, y^{j} \in [0,1].
\end{aligned}
\end{equation}
$y^{j}$=1 denotes the $j$-th word appearing in the input and vice versa.

\subsubsection{Contrastive Loss}

The contrastive loss is applied on word distributions $\tilde{\textbf{z}}$ which is normalized from $\hat{z}$ in Eq.~(\ref{alg:sigmoid}).
Two word distributions of masked versions of one text are pulled close while pushing away from other word distributions.
We use Jensen–Shannon divergence function (JS) \cite{Lin1991DivergenceMB} to compute the similarity between word distributions.

Given a mini-batch, the contrastive loss over $2m$ masked sequences is defined as follows, 
\begin{equation}\label{ctl}
\begin{aligned}
\mathcal{L}_{CL} = -\sum_{i=1}^{m} \log \frac{exp(-JS(\tilde{\textbf{z}}_{i}^{'},\tilde{\textbf{z}}_{i}^{''}))}{\sum_{k=1}^{2m} \mathbbm{1}_{[k\neq{i}]} exp(-JS(\tilde{\textbf{z}}_i,\tilde{\textbf{z}}_k))},
\end{aligned}
\end{equation}
where $(\tilde{\textbf{z}}_{i}^{'},\tilde{\textbf{z}}_{i}^{''}))$ are two masked version of one text $d_i$.
We also propose a variant which directly contrasts the dense representations $\textbf{h}$, and this variant is denoted as CPDAE$_R$.

\subsubsection{MLM Loss}
Similar to existing works \cite{lu-etal-2021-seed,Ma2021BPROPBP}, we also  adopt the MLM \cite{Devlin2019BERT} to build good word representations. 
We omit its details here and refer the reader to the original BERT paper~\cite{Devlin2019BERT}.

The final loss is the total sum of MLM loss, reconstruction loss and contrastive loss, which is formulated as, 
\begin{equation}\label{total-loss}
\begin{aligned}
\mathcal{L}_{total} = \mathcal{L}_{REC} + \mathcal{L}_{MLM} + \lambda{\mathcal{L}_{CL}},
\end{aligned}
\end{equation}
where $\lambda$ is a hyper-parameter.

\subsection{Theoretical Analysis}\label{theory}
We mathematically show why our contrastive pre-training based on word distributions can learn a discriminative autoencoder.

A natural language text is composed of a large portion of common words and a small portion of representative/informative words.
Suppose $S \subset V$ denote the common words in the $V$ and $\complement{V}S $ denote the rest words.
According to Equation~(\ref{ctl}), $\mathcal{L}_{CL}$ aims to minimize $-JS(P, Q)$ between word distributions from different input texts, and maximize $-JS(P, Q)$ between word distributions from the same input texts.
The $-JS(P, Q)$ can be rewritten to the following:
\begin{equation}\label{eq:common}
\begin{aligned}
-JS(P, Q) =& -\sum_{x \in |V|} p(x)log(p(x))- \sum_{x \in |V|} q(x)log(q(x)) \\ 
+& \sum_{x \in |V|} (p(x)+q(x))log(p(x)+q(x)). \\
\end{aligned}
\end{equation}
where we ignore $log2$ as it is a constant.

We will discuss two situations, i.e., word $x \in S$ and word $x \in \complement{V}S $.
For the word $x \in S$, $p(x)$ is equal to $q(x)$ as the common words appear in almost every document with the same high probabilities $a$.
Thus, Equation~(\ref{eq:common}) can be reduced to:
\begin{equation}\label{the}
\begin{aligned}
-JS(P, Q)_{x \in S} =& -\sum_{x \in S} 2p(x)log(p(x)) + \sum_{x \in S} 2p(x)log(2p(x)) \\
=& \sum_{x \in S} 2p(x)log2.
\end{aligned}
\end{equation}
Therefore, given a mini-batch of $2m$ examples, $p(x)$ will be lowered as there are $2m-2$ word distributions needed to be minimized and only 1 positive word distributions needed to be maximized.
So we conclude that (1) \textbf{word distribution based contrastive pre-training will suppress the probability of common words when decoding}.

For the word $x \in \complement{V}S$, the contrastive loss needs maximize Eq.~(\ref{the}) between two word distributions from the same text. 
For word distributions from other texts,  $q(x)$ is close to 0 since representative words only occur in $p(x)$.
The contrastive loss needs minimize Equation~(\ref{eq:common}) which can be reduced to,
\begin{equation}\label{rep}
\begin{aligned}
 -JS(P, Q) = 0. \\
\end{aligned}
\end{equation}
In summary,
(2) \textbf{word distribution based contrastive pre-training  will highlight the probability of representative/informative words when decoding}.

\begin{table*}[t]
  \renewcommand{\arraystretch}{0.75}
  \caption{Comparisons between CPDAE and the baselines. Two-tailed t-tests demonstrate the improvements of CPDAE to the baselines are statistically significant ($p \le 0.05$). $\ast, \dag, \ddag$ indicates significant improvements over BERT, ICT, and SEED, respectively. Results not available or not applicable are marked as `-'.}
  \label{tab:main_res}
  \begin{tabular}{lllllllll}
  \toprule
     \multirow{2}{*}{Model} & \multicolumn{2}{c}{MARCO Dev Passage} & \multicolumn{2}{c}{TREC2019 Passage} & \multicolumn{2}{c}{MARCO Dev Doc} & \multicolumn{2}{c}{TREC2019 Doc}  \\
    \cmidrule(lr){2-3} \cmidrule(lr){4-5} \cmidrule(lr){6-7} \cmidrule(lr){8-9}
     &  MRR@10 & Recall@1000 & NDCG@10 & Recall@1000 & MRR@100 & Recall@100  & NDCG@10 & Recall@100  \\ 
    \midrule
    BM25 & 0.187 & 0.857 & 0.501 & 0.745 & 0.277 & 0.808 & 0.519 & \textbf{0.395}\\
    Best TREC Trad\cite{Craswell2020OverviewOT} & - & - & 0.554 & - & - & - & 0.549 & -\\
    \midrule
    BERT & 0.335 &  0.957 & 0.661 & 0.769  & 0.389 & 0.877 & 0.594 & 0.301 \\
    SimCSE & 0.335 & 0.955 & 0.662 & 0.766 & 0.391 & 0.879 & 0.598 & 0.302\\
    ICT & 0.339 & 0.955 & 0.670 & 0.775 & 0.396 & 0.882 & 0.605 & 0.303\\
    PROP & 0.337 & 0.951 & 0.673 & 0.771 & 0.394 & 0.884 & 0.596 & 0.298 \\
    SEED & 0.342$^{\ast}$ & 0.963 & 0.679$^{\ast}$ & 0.782$^{\ast\dag}$ & 0.396 & 0.902$^{\ast}$ & 0.605$^{\ast}$ & 0.307  \\
    \midrule
    CPDAE$_{R}$ & 0.350$^{\ast\dag}$ & 0.965$^{\ast\dag}$ & 0.686$^{\ast\dag}$ & 0.789$^{\ast\dag}$ & 0.402$^{\ast}$ & \textbf{0.909}$^{\ast\dag}$ & 0.609$^{\ast}$ & 0.311$^{\ast}$ \\
    CPDAE & \textbf{0.355}$^{\ast\dag\ddag}$ & \textbf{0.968}$^{\ast\dag}$ & \textbf{0.696}$^{\ast\dag\ddag}$ & \textbf{0.799}$^{\ast\dag\ddag}$ & \textbf{0.408}$^{\ast\dag\ddag}$ & 0.907$^{\ast\dag}$ & \textbf{0.615}$^{\ast\dag\ddag}$ & 0.315$^{\ast\dag}$ \\
    \bottomrule
  \end{tabular}
\end{table*}

\section{Experiments}
In this section, we conduct experiments to demonstrate the effectiveness of our proposed model.

\subsection{Experimental Settings}

Here, we introduce the pre-training corpus, downstream tasks, baseline methods, and implementation details.

\subsubsection{Pre-training Corpus and Downstream Tasks}
We use the English Wikipedia as our pre-training corpus following previous works~\cite{Devlin2019BERT,lu-etal-2021-seed,Chang2020PretrainingTF,Ma2021PROPPW,Ma2021BPROPBP}.
We conduct experiments on several public dense retrieval benchmarks, including MS MARCO Passage Ranking (MARCO Dev Passage)~\cite{Campos2016MSMA}, MS MARCO Document Ranking (MARCO Dev Document)~\cite{Campos2016MSMA}, TREC 2019 Passage Ranking (TREC2019 Passage)~\cite{Craswell2020OverviewOT} and TREC 2019 Document Ranking (TREC2019 Document)~\cite{Craswell2020OverviewOT}.

\subsubsection{Baselines}
We adopt the traditional sparse retrieval models and pre-trained models as baselines.  
For traditional spare retrieval models, we choose the strong \textbf{BM25} method \cite{Robertson2009ThePR}. 
We also list several representative results according to the TREC overview paper~\cite{Craswell2020OverviewOT}. 
For the pre-trained models, besides \textbf{BERT}~\cite{Devlin2019BERT}, the main baseline is the state-of-the-art autoencoder-based language models \textbf{SEED}  \cite{lu-etal-2021-seed}.
We also consider two contrastive learning methods, including \textbf{ICT}~\cite{Chang2020PretrainingTF}, and \textbf{SimCSE}~\cite{Gao2021SimCSESC}.
We also consider the state-of-the-art pre-trained models on the re-ranking, i.e., \textbf{PROP}~\cite{Ma2021PROPPW}, but we pre-train it with a bi-encoder for a fair comparison.


\subsubsection{Implementation Details}

\begin{itemize}[leftmargin=*]
\item \textbf{Pre-training details}.
We use BERT to initialize our encoder. 
The output hidden size of the MLP decoder is set to 30522 which is the size of BERT's vocabulary.
We use a learning rate of 5e-5 and Adam optimizer with a linear warm-up technique over the first 10\% steps.
We pre-train on Wikipedia for 3 epochs with the batch size $m$ as 64. 
$\lambda$ is set to 0.1 in the Eq.~(\ref{total-loss}).

\item \textbf{Fine-tuning details}.
The decoder is only used in pre-training and is dropped during fine-tuning. 
Following previous works~\cite{Xiong2021ApproximateNN,Zhan2021OptimizingDR,lu-etal-2021-seed,Gao2021CondenserAP}, we employ a bi-encoder architecture based on the encoder of CPDAE and use a pairwise loss for fine-tuning. 
We use a learning rate of 5e-6, a batch size of 64, and pair each positive example with 7 negative examples. 
For two passage ranking datasets (i.e., MACRO Dev Passage and TREC2019 Passage), we train the model with static hard negatives using the BM25 warm-up model following~\cite{Zhan2021OptimizingDR}.
For two document ranking datasets (i.e., MARCO Dev Doc and TREC2019 Doc), we use the model fine-tuned on the passage ranking task as the starting point following \cite{Xiong2021ApproximateNN, Zhan2021OptimizingDR,lu-etal-2021-seed}. 
We then iteratively mine the static hard negatives using the current model twice and fine-tune the model for 1 epoch in each iteration.

\end{itemize}

\subsection{Baseline Comparison}\label{sec:main_results}

The performance comparisons between CPDAE and the baselines are shown in Table \ref{tab:main_res}. We have the following observations: 
(1) Dense retrieval models generally outperform the traditional sparse retrieval models by a large margin on most of the datasets. 
This is mainly because dense retrieval models could well capture the semantics meanings of queries and documents and can over the vocabulary mismatch problem. 
(2) SimCSE, ICT and PROP show slight improvements over BERT, indicating that these pre-training methods may not be optimal for dense retrieval. 
ICT only pulls a random sentence close to its context in the representation space, while the random sentence may be semantic similar to other texts and thus not be distinguishable from different texts.
(4) SEED performs the best among all the baseline, indicating the autoencoder-based language models can produce high-quality dense representation via reconstruction.

We find that CPDAE can generally outperform baseline methods significantly, including general pre-trained language model BERT, other contrastive learning methods, and autoencoder-based language models. 
The better results demonstrate the effectiveness of our contrastive loss to encode discriminative text sequence representations.  
CPDAE performs better than CPDAE$_{R}$ even though contrasting dense representations in the representation space is more straightforward.
But the performance of contrasting dense representations heavily depends on the data augmentation while we only use a weak randomly masking.

\begin{table}[t]
    \caption{Ablation studies of the contrastive loss (CL) in CPDAE. IDF-REC: using IDF to weight reconstruction loss. Best results are marked bold.}
\renewcommand{\arraystretch}{0.9}
    \setlength\tabcolsep{6pt}
    \centering
    \begin{tabular}{lccccc}
    \toprule
      \multirow{2}{*}{} & \multicolumn{2}{c}{MARCO Dev Passage} && \multicolumn{2}{c}{MARCO Dev Doc} \\
    \cline{2-3} \cline{5-6}
    & MRR@10 & R@1000 && MRR@100 & R@100
    \\
    \midrule
        SEED & 0.342 & 0.963 && 0.396 & 0.902  \\
        w/o CL & 0.344 & 0.963 && 0.397 & 0.905  \\
        IDF-REC & 0.347  & 0.962 && 0.399 & 0.906 \\
        w/ CL & \textbf{0.355} & \textbf{0.968} && \textbf{0.408} & \textbf{0.907} \\
        \bottomrule
    \end{tabular}

    \label{tab:cl}
\end{table}

\begin{table}[t]
    \caption{Fine-tuning with limited supervised data. Performance is measured by Recall@1000 and Recall@100 for MARCO Dev Passage and MARCO Dev Doc, respectively.}
\renewcommand{\arraystretch}{0.9}
    \setlength\tabcolsep{5pt}
    \centering
    \begin{tabular}{lccccccc}
    \toprule
      \multirow{2}{*}{} & \multicolumn{3}{c}{MARCO Dev Passage} && \multicolumn{3}{c}{MARCO Dev Doc} \\
    \cline{2-4} \cline{6-8}
    & 0.1k & 1k & 10k & & 0.1k & 1k & 10k
    \\
    \midrule
        BERT & 0.636 & 0.803 & 0.891 && 0.512 & 0.692 & 0.784 \\
        SEED & 0.659 & 0.827 & 0.914 && 0.523 & 0.717 & 0.835 \\
        CPDAE & \textbf{0.708}  & \textbf{0.855} & \textbf{0.923} && \textbf{0.573} & \textbf{0.821} & \textbf{0.868} \\
        \bottomrule
    \end{tabular}
    \label{tab:low-res}
\end{table}

\subsection{Ablation Analysis}

We conduct an ablation analysis to investigate the effect of the proposed contrastive loss (CL) in our CPDAE. 
We also compare with a weighted reconstruction loss (IDF-REC) which weights each token loss with IDF value in Eq.~(\ref{recs}).
As shown in Table~\ref{tab:cl}, we can find that: 
(1) By removing the CL, the performance of CPDAE (w/o CL) is slightly better than SEED, indicating the effectiveness of the novel reconstruction loss with a non-autoregressive decoder by avoiding the bypass effect. 
(2) IDF-REC and CPDAE (w/o CL) have a similar performance while both perform significantly worse than CPDAE (w CL), again demonstrating the proposed contrastive loss could help learn discriminative representations.

\subsection{Low-Resource Setting and Visual Analysis}

To further illustrate the effectiveness of CPDAE, we simulate a low-resource setting on the MARCO Dev Passage and MARCO Dev Doc respectively. 
We randomly sample a limit number of queries (i.e., 0.1k, 1k, 10k) from the original training set to fine-tune the pre-trained models. 
Each experimental result is reported as the average of three runs with different sampled queries. 
As shown in Table~\ref{tab:low-res}, we can see that CPDAE outperforms BERT and SEED on all the datasets using the same number of limited supervised data.  
This result demonstrates that CPDAE can provide more discriminative text representations than BERT and SEED.

\begin{table}[t]
    \caption{Visualization of the word distributions generated by the MLP decoder in CPDAE. Darker color indicates higher probability.}
    \small
    \renewcommand{\arraystretch}{0.25}
    \centering
    \begin{tabular}{p{8.2cm}}
    \toprule
    \multicolumn{1}{l}{Probability:  \colorbox{red!0}{\makebox(2,3){0}},  \colorbox{red!10}{\makebox(8,4){0.1}}, \colorbox{red!30}{\makebox(8,4){0.2}}, \colorbox{red!40}{\makebox(8,4){0.3}}, \colorbox{red!50}{\makebox(8,4){0.4}},
    \colorbox{red!70}{\makebox(10,4){>0.5}}} \\
    \midrule
    \colorbox{red!40}{ Nanotechnology}  \colorbox{red!2}{raises} many of the same  \colorbox{red!3}{issues} as any new  \colorbox{red!10}{technology}, including  \colorbox{red!3}{concerns} about the  \colorbox{red!10}{toxicity} and  \colorbox{red!10}{environmental} impact of  \colorbox{red!10}{nanomaterials}, and their  \colorbox{red!5}{potential}  \colorbox{red!3}{effects} on global  \colorbox{red!5}{economics}, as well as \colorbox{red!10}{speculation} ... \\

\bottomrule
\end{tabular}
\label{tab:vis1}
\end{table}



To illustrate how CPDAE improves retrieval performance, we visualize the normalized word distributions generated by the MLP decoder in Table~\ref{tab:vis1}. 
We randomly sample a short piece of text from Wikipedia and sum up the normalized output probabilities of all the subwords of a whole word.  
As shown in Table~\ref{tab:vis1},  we can see that CPDAE can suppress the common words and highlight the informative words as shown in the theoretical analysis~\ref{theory}.


\section{Conclusion}
In this paper, we present a contrastive pre-training method to learn a discriminative autoencoder for dense retrieval. 
We propose to employ a non-autoregressive MLP decoder to avoid the bypass effect and apply contrastive learning to the word distributions produced by the decoder. 
We theoretically show that our contrastive strategy can suppress the common words and highlight the representative words, leading to discriminative representations.
Experiments at four public dense retrieval benchmarks show that our method could achieve significant improvements over the baselines.
For future work, we would like to investigate the data augmentation techniques and apply our method to other IR scenarios.

\begin{acks}
This work was funded by the National Natural Science Foundation of China (NSFC) under Grants No. 62006218 and 61902381, the Youth Innovation Promotion Association CAS under Grants No. 20144310, and 2021100, the Young Elite Scientist Sponsorship
Program by CAST under Grants No. YESS20200121, and the Lenovo-CAS Joint Lab Youth Scientist Project.
\end{acks}

\newpage
\bibliographystyle{ACM-Reference-Format}
\balance
\bibliography{main}



\end{document}